# Time-domain impedance method for transient photovoltage analysis

A. Nadtochiy [1], A. Podolian [1], O. Korotchenkov[1,2]

[1] *Faculty of Physics, Taras Shevchenko National University of Kyiv, Kyiv 01601, Ukraine*
[2] *Erwin Schrödinger International Institute for Mathematics and Physics, University of Vienna, Vienna 1090, Austria*

In this work, we approximate the surface photovoltage (SPV) transients in nm-sized ZnO films by the equivalent RC circuit model. The SPV rises in time during ≈90 μs after the exciting light pulse at 275 nm is off at different pulse widths ranging from 1.2 to 12 μs. The key to this observation is a considerable amount of defects in the films, which form a trap capacitance in the equivalent circuit. The photogeneration of nonequilibrium electrons and holes near the film surface is described by charging of a capacitance by the current source whereas the rate of their spatial separation is determined by a resistance. This resistance reflects an obstacle in the carrier movement while another capacitance determines the charge separation distance. The electron-hole recombination is account for a second resistance introduced into the equivalent circuit. The resulting modeled SPV transient allows to reproduce the observed experimental curve rather well.

## 1. INTRODUCTION

In order to gain an understanding of the interfacial electronic properties and the band bending at the substrate/layer interface, the metal–insulator (oxide)–semiconductor capacitance and conductance upon varying gate voltage and frequency can generally be measured [1]. Analysis of the frequency-dependent capacitance and conductance uses an equivalent circuit, which can take into account effects of interface states. It is also possible to employ the impedance spectroscopy technique capturing an electrical response of the structure to the ac signal in a broad frequency range [2]. This approach is focused on searching of the simplest equivalent electrical circuit, including resistance ($R$), capacitance ($C$) inductance ($L$) elements, that reproduces the impedance of the measured structure. CPE was originally used in electrochemistry to characterize solid-electrolyte systems. Additional elements, e.g. the constant phase element (CPE), can be supplemented to address a nonideal capacitor behavior, particularly in electrochemical impedance spectroscopy [3]. The equivalent circuits with CPEs were also used to tackle the problem of surface roughness, leakage capacitance and nonuniform charge distribution in semiconductor-insulator systems that result in the nonideal capacitor behavior [1,4–7].

Conditions for charge injection at the substrate/layer interface may significantly affect the dynamics of charge transport and separation across the space charge layer. These dynamics can be probed utilizing the surface photovoltage (SPV) decays.

All the above mentioned consequential effects should, in principle, be taken into account in the equivalent circuit computations, thereby making it truly difficult to get a suitable model. The use of the frequency- and time-domain impedance measurements for the parametrization of equivalent circuit models is widely applicable [8–13]. The transmission line model (TLM) and the full wave model (FWM) are the most commonly used methods for quantifying transient impedance [12].

Here, we realize the time-domain impedance analysis to fit the experimentally observed SPV decay curves.

## 2. EXPERIMENTAL

Experimental SPV decay curves were taken in ZnO nm-sized films, which were deposited onto Si(100) substrates at room temperature by the magnetron sputtering using a ZnO (99.99% purity) target with Al content of 2%. Details of the growth procedure were previously discussed [14]. SPV transients were measured in the capacitor arrangement [15], and details of our setup were given elsewhere [16].

## 3. RESULTS AND DISCUSSION

The SPV transient curve taken in a 70 nm ZnO film is shown in Fig. 1.

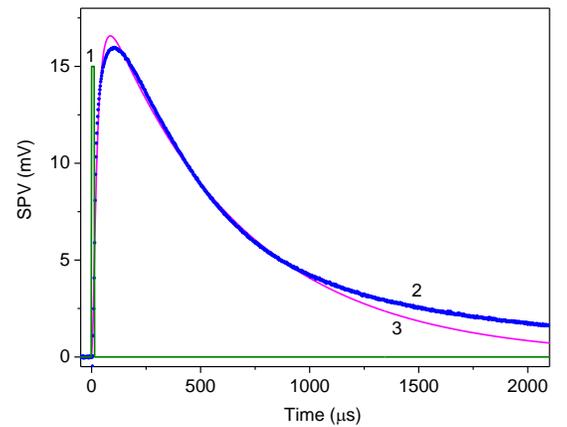

FIG. 1. SPV transients in ZnO film: 1 – exciting light pulse with the duration of 12 μs, 2 – experimental SPV transient, 3 – fitting curve for the circuit of Fig. 2 at $C_1$ = 2.1 nF, $R_1$ = 11 kΩ, $C_2$ = 1.12 μF, $R_2$ = 564 Ω.

Schematic diagram of the equivalent circuit of the SPV measurement is shown in Fig. 2. The SPV decay is formed by the $C_1$, $R_1$, $C_2$, and $R_2$ elements. The other circuit elements, $R_3 - R_5$ and $C_3 - C_4$, belong to the SPV signal amplification scheme, which was considered elsewhere [17].

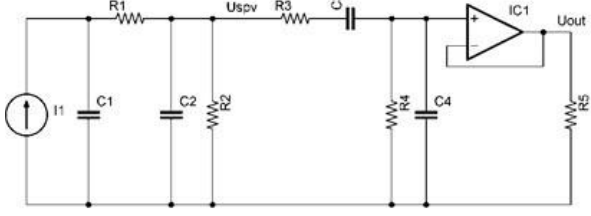

FIG. 2. Equivalent circuit used to fit the SPV transient in Fig. 1. $I_1$ – photo-generated current, $U_{SPV}$ is the generated SPV signal.

Importantly, the experimental transient in Fig. 1 (curve 2) is growing some time after the light was turned off (curve 1 in Fig. 1), reaches a maximum at about 100 μs, and then starts to decrease rapidly. In a monoexponential approximation, the characteristic relaxation time $\tau$ of the decreasing part of the curve can be found from the dependence $U_{SPV} = A\, exp(-t/\tau)$. One gets $\tau = 633$ μs, which is very close to the value of $\tau = C_2 R_2 = \cdot 632$ μs. Therefore, if $C_1$ and $R_1$ are excluded from the circuit, the $C_2$ and $R_2$ elements will describe the ordinary SPV decays. In turn, the $C_1$ and $R_1$ elements form a delay circuit such that, during the light pulse, the energy is firstly accumulated in $C_1$ and then flows into the $C_2$ and $R_2$ elements through $R_1$. The $C_2$ and $R_2$ elements thus form the main channel of the relaxed SPV.

Let's consider in more detail the method of approximating experimental data using the fitting of $C_1$, $R_1$, $C_2$, and $R_2$. This process consists of two steps. At the first step, it is necessary to calculate the voltage $U_{SPV}$ depending on the applied current $I_1$. At the second step, we minimize the difference between the experimental data and the calculation by varying the five parameters $I_1$, $C_1$, $R_1$, $C_2$, and $R_2$.

The solution of the first step is based on the theory of electrical circuits [18]. Using the Kirchhoff's laws one can write the current versus voltage equations for the resistance and capacitance as
$$u_R = I_R R,$$
$$I_C = C\frac{du_C}{dt}.$$

We then obtain a system of ordinary differential equations in the form

$$\begin{bmatrix}\frac{du_{C1}}{dt}\\ \frac{du_{C2}}{dt}\end{bmatrix} = \begin{bmatrix} -\frac{1}{C_1 R_1} & \frac{1}{C_1 R_1} \\ \frac{1}{C_2 R_1} & -\frac{R_1+R_2}{C_2 R_1 R_2}\end{bmatrix} \cdot \begin{bmatrix}u_{C1}\\u_{C2}\end{bmatrix} + \begin{bmatrix}\frac{I_1}{C_1}\\0\end{bmatrix}.$$

Here, $I_1$ depends on time and the light pulse amplitude $I_1 = A_{pulse} e(t)$, $u_{C1}$ and $u_{C2}$ are the voltages on the capacitors $C_1$ and $C_2$.

The above system of equations was solved using the Backward Euler method [18] to get the voltage $U_{SPV} = u_{C2}$. Then, the $I_1$, $C_1$, $R_1$, $C_2$, and $R_2$ parameters were varied to minimize the difference between the experimental and calculated curves of the time-dependent $U_{SPV}$. This procedure was performed using the NLopt program library [19].

Figure 1 shows that the calculated value of $U_{SPV}$ describes the initial part of the SPV decay curve quite well, but, at latter times, there is a divergence from the experimental curve. This is clearly because the SPV decay is not monoexponential, but consists of several exponents. In order to describe the presence of multi-exponential contributions in the decay curves, it is necessary to modify the equivalent scheme given in Fig. 2. Thus, Fig. 3 exemplifies an equivalent circuit for a two-exponential dependence.

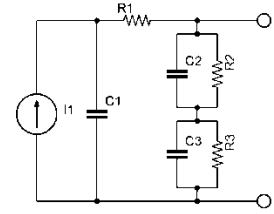

FIG. 3. Equivalent circuit modified to account the two-exponential SPV decay. The circuit elements $C_2$, $R_2$, $C_3$, and $R_3$ are responsible for the formation of the two-exponential behavior.

Let's define $u_{C1}$, $u_{C2}$, and $u_{C3}$ to be the voltages on the capacitors $C_1$, $C_2$, and $C_3$, respectively. Then the system of ordinary differential equations describing the circuit in Fig. 3 takes the form

$$\begin{bmatrix}\frac{du_{C1}}{dt}\\ \frac{du_{C2}}{dt}\\ \frac{du_{C3}}{dt}\end{bmatrix} = \begin{bmatrix} -\frac{1}{C_1 R_1} & \frac{1}{C_1 R_1} & \frac{1}{C_1 R_1}\\ \frac{1}{C_2 R_1} & -\frac{R_1+R_2}{C_2 R_1 R_2} & -\frac{1}{C_2 R_1}\\ \frac{1}{C_3 R_1} & \frac{1}{C_3 R_1} & -\frac{R_1+R_3}{C_3 R_1 R_3}\end{bmatrix} \cdot \begin{bmatrix}u_{C1}\\u_{C2}\\u_{C3}\end{bmatrix}$$
$$+\begin{bmatrix}\frac{I_1}{C_1}\\0\\0\end{bmatrix}.$$

After performing all the above-described procedures for optimizing the circuit parameters $I_1$, $I_1$, $C_1$, $R_1$, $C_2$, $R_2$, $C_3$, and $R_3$, one obtains the curve of $U_{SPV} = u_{C2} + u_{C3}$, which is plotted by line 4 in Fig. 4. Evidently, curve 4 produces much better agreement with the experimental data (curve 2) than curve 3.

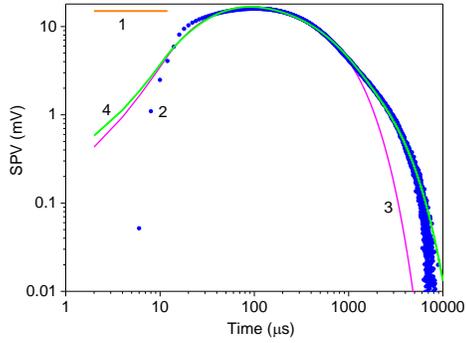

Fig. 4. SPV transients in ZnO film. Line 1 identifies the position and duration of the exciting light pulse, 2 – experimental SPV transient (curve 2 in Fig. 1 replotted in the double-log scale), 3 – fitting curve for the circuit of Fig. 2 (curve 3 in Fig. 1), 4 – fitting curve for the circuit of Fig. 3.

Varying the light pulse length from 1.2 μs and 12 μs (Fig. 3) yields quite similar SPV transient shapes, as evidenced in Fig. 5. The SPV signal reaches a maximum at essentially the same time moment of about 90 μs in both cases, calculated from the beginning of the light pulses.

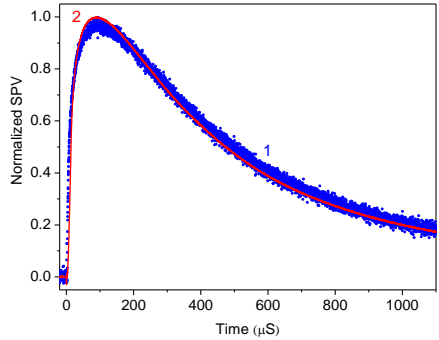

Fig. 5. SPV transients in ZnO film excited by light pulses with the duration of 1.2 (1) and 12 (2) μs. The curves are normalized to unity.

It is known that formation of the SPV signal occurs as a result of spatial separation of photogenerated charges. The band gap of ZnO is 3.37 eV, which corresponds to the light wavelength of 368 nm. Therefore, the LED light with a wavelength of 275 nm used here is strongly absorbed within the ZnO layer, which in turn causes a very inhomogeneous distribution of photogenerated carriers across the layer thickness.

Assuming that the spatial separation of the carriers occur only due to their diffusion, one expects that the SPV signal would reach its maximum value at different points in time, depending on the duration of the exciting light pulse. However, due to large defect densities in the layer, the capture of charge carriers at the defect levels would affect the SPV transient shape. Therefore, the data of Fig. 5 indicate that the capture processes and a subsequent slow release of the captured carriers from the defect levels limit the rate of the spatial separation of charge carriers.

It is possible to relate the equivalent circuit elements to the physical processes occurring in the structure. Thus, the charging of the capacitance $C_1$ by the current source $I_1$ corresponds to the photogeneration of non-equilibrium electrons and holes near the surface of the ZnO film. The rate of spatial separation of photogenerated carriers and, accordingly, the growth rate of the SPV signal is determined by $R_1$, which acts as an obstacle in the spatial separation of charge carriers. The greater the value of $R_1$, the slower the carriers move.

The capacitance $C_2$ is formed due to the spatial separation of the negatively charged electrons and positively charged holes. The value of the capacitance $C_2$ determines the separation distance of the photogenerated carriers, and the voltage value to which $C_2$ is charged is determined by the number of photogenerated carriers.

Since after turning off the light pulse, the carrier recombination occur in parallel with the spatial separation of the charge carriers, the $R_2$ resistance is introduced into the equivalent circuit. It causes the discharge of $C_2$ due to the recombination processes. The value of $R_2$ therefore determines the recombination rate. The faster recombination rate is suggested by the smaller $R_2$ value.

In this picture, introducing additional capture centers can be accounted for by introducing additional $RC$ elements into the equivalent circuit used to model the SPV transient shape.

## 4. CONCLUSIONS

In summary, the SPV transients in nm-sized ZnO films can be described by the equivalent RC circuit model. The photogeneration of nonequilibrium electrons and holes near the film surface is described by charging of the capacitance $C_1$ by the current source $I_1$. The rate of spatial separation of the electrons and holes is determined by the $R_1$ resistance, which reflects an obstacle in the spatial separation of charge carriers. The capacitance $C_2$ determines the carrier separation distance. The $R_2$ resistance is introduced into the equivalent circuit to account for the discharge of $C_2$ due to electron-hole recombination processes. Faster recombination rates make $R_2$ smaller. The resulting modeled SPV transients reproduce the experimental data rather well.

**Acknowledgments**
This research was supported by the Ministry of Education and Science of Ukraine, grant number 0122U001953. O.K. acknowledges support from the Erwin Schrödinger Institute by the Special Research Fellow Programme.